\documentstyle[12pt,aaspp4]{article}

 \mathcode`*="002A   
\def\lta{\mathbin{\lower 3pt\hbox
     {$\rlap{\raise 5pt\hbox{$\char'074$}}\mathchar"7218$}}}   
\def\gta{\mathbin{\lower 3pt\hbox
     {$\rlap{\raise 5pt\hbox{$\char'076$}}\mathchar"7218$}}}   

\mathchardef\star="313F

\def\Re{\mathop{\it Re}\nolimits}

\def\den{\tilde \rho_1}
\def\energy{\tilde E_1}
\def\fcor{\tilde f_{{\rm co}\,r1}}
\def\Fcor{\tilde F_{{\rm co}\,r1}}
\def\fphi{\tilde F_{\phi 1}}
\def\fr{\tilde F_{\rm r1}}
\def\ftheta{\tilde F_{\theta 1}}
\def\fphi{\tilde F_{\phi 1}}
\def\fthetaco{\tilde f_{{\rm co\,}\theta 1}}
\def\grr{\left(1-r_s/r\right)^{-1}}
\def\ur{\tilde u^r_1}
\def\ut{\tilde u^t_1}
\def\utheta{\tilde u^{\hat\theta}_1}
\def\uphi{\tilde u^{\hat\phi}_1}
\def\utzero{u_0^t}

\begin{document}

\title{Variability and Stability in\\Radiation Hydrodynamic Accretion Flows}
\author{Guy S. Miller\altaffilmark{1}}
\affil{Department of Physics and Astronomy\\Northwestern University\\
Evanston, IL 60208, USA}
\and
\author{Myeong-Gu Park\altaffilmark{2}}
\affil{Department of Astronomy and Atmospheric Sciences\\
Kyungpook National University\\
Taegu 702-701, Korea}
\altaffiltext{1}{E-mail: gsmiller@casbah.acns.nwu.edu}
\altaffiltext{2}{E-mail: mgp@kyungpook.ac.kr}

\slugcomment{Submitted to {\it The Astrophysical Journal}}

\begin{abstract}

In this paper we examine time-dependent and three-dimensional perturbations of
spherical accretion flow onto a neutron star close to its Eddington limit.  Our
treatment assumes a Schwarzschild geometry for the spacetime outside the neutron
star and is fully general relativistic.  At all the accretion rates studied,
the response of the accretion flow to perturbations includes weakly damped
oscillatory modes.  At sufficiently high luminosities --- but still well below
the Eddington limit --- the flows become unstable to aspherical perturbations. 
These unstable radiation hydrodynamic modes resemble the onset of convection,
and allow accretion to occur preferentially through more rapidly descending
columns of gas, while the radiation produced escapes through neighboring
columns in which the gas descends more slowly.

\end{abstract}


\newpage

\section{Introduction}

Accreting neutron stars and black hole candidates exhibit a rich variety of
temporal variabilities.  In addition to X-ray pulsations, flares, bursts, and
eclipses, quasiperiodic oscillations (QPO) are frequently observed.  These are
almost periodic variations in the radiation from a source, concentrated in
relatively narrow frequency intervals, so that frequencies for the QPO can be
meaningfully defined.
Several types of QPO have been identified in black hole candidate systems and
in systems believed to contain accreting neutron stars.  Each has its own
distinctive phenomenology, and presumably each has a different underlying
mechanism.
Observations have shown that changes in the X-ray spectrum of a source are
accompanied by corresponding changes in the frequencies, strengths, and widths
of its QPO (for reviews, see Lewin, van Paradijs, \& van der Klis~1988; and van
der Klis 1989, 1993, and~1995).  It is thus likely that a direct connection exists
between the QPO and the production of radiation in the central engines of these
sources.  The development of models for QPO has taken a place on the agenda of
high-energy astrophysics comparable in importance to the modeling of spectra:
both are needed to obtain a clearer picture of how accreting systems work.

In searching for guides to the interpretation of quasiperiodic oscillations, it
is helpful to examine the simplest possible models and ask what temporal
variability they might engender. A classic paradigm for the study of accreting
systems is spherical accretion. Although spherical accretion is obviously a
gross idealization of naturally occurring astrophysical flows, it remains a
useful preliminary to the study of more realistic, and correspondingly more
complex systems.  Studies of spherical accretion also may be directly
applicable to components of an accretion flow with little angular momentum;
e.g., if the accretion flow consists of a disk flow embedded in a coronal
inflow, the coronal component might be treated as spherical to a first
approximation (Lamb 1989; Fortner, Lamb, \&~Miller~1989, hereafter FLM89).

The temporal variability of the brightest accretion-powered sources may be a
consequence of radiation hydrodynamics.  Steady spherical accretion becomes
impossible when the luminosity of the accreting system exceeds the Eddington
limit $L_E$.  At the Eddington limit, the force on gas due to radiation
escaping from the system exactly counterbalances the gravitational attraction
of the central star.  At higher luminosities, accretion can proceed only
aspherically (see, e.g., the thick accretion disk models of Abramowicz,
Calvani, \& Nobili~1980) or sporadically.

Many studies exist of steady, spherical radiation-hydrodynamic flow.  Here we
limit ourselves to the case of accretion by a neutron star (references to the
literature on steady spherical black hole accretion are collected in Zampieri,
Miller, \& Turolla~1996, which contains a recent treatment of time-dependent
spherical flow onto a black hole).  Early numerical work by Shakura~(1974),
by Maraschi, Reina, and Treves (1974,~1978), and by Vitello~(1978) explored the
interaction of a steady spherical flow with the radiation it produces, treating
the gravitational field of the accreting star Newtonianly.  Analytic
approximations for steady Newtonian flow solutions were obtained by
Miller~(1990).  The Newtonian approximation was dropped by Park and
Miller~(1991), who obtained numerical solutions for steady accretion onto a
neutron star with a fully general relativistic treatment.  They found that the
``settling regime'' of the accretion flow, where the infalling gas decelerates
and settles onto the stellar surface, is severely modified by the effects of
general relativity: general relativistic flows tend to decelerate to much lower
speeds than Newtonian flows, and hence have far greater densities and optical
depths.  Although one might anticipate that general relativistic modifications
of flow solutions would be small (of the order of the dimensionless
gravitational potential at the neutron star surface, $\sim 2GM/[Rc^2]\approx
0.2$), the near stalemate between radiation forces and gravitation in the
settling regime makes general relativity disproportionately important.

Several authors have studied time-dependent spherical flow onto a neutron star.
Klein, Stockman, \& Chevalier~(1980) performed numerical simulations of
supercritical accretion, in which mass was supplied to the neutron star at
rates $\dot M$ much greater than the Eddington rate $\dot M_E$  necessary to
produce a luminosity $L_E$.  They found that their flows became choked with
radiation, so that accretion in the inner parts of the flow proceeded only fast
enough to bring the system luminosity up to the Eddington limit.  The same
conclusions were reached by Burger \& Katz~(1983), who conducted a similar
study.  Both groups treated the neutron star's gravitational field in the
Newtonian approximation, and followed the development of their flows for
relatively short times, typically tens of milliseconds.  Fortner, Lamb, \&
Miller~(1989) examined spherical accretion at rates $\dot M$ slightly smaller
than the critical rate $\dot M_E$, and allowed their flows to evolve for much
longer times.  They found that near-critical flows can develop global
oscillations, in which episodes of stronger inflow alternate with periods
during which the radiation escapes.  All of the flows in these investigations
were restricted to strict spherical symmetry.

Departures from spherical symmetry allow radiation to escape more readily
through some regions of the flow, while material accretes preferentially
elsewhere.  Aspherical flows were studied by Miller and Park (1995; hereafter
MP95).  They treated the problem perturbatively by linearizing the radiation
hydrodynamic equations about steady, spherical flow solutions; thus, fully
asymmetrical flows could be examined, as long as departures from the
underlying spherical flow solutions were not too great.  The resulting
quasispherical flow solutions included both the spherical oscillatory modes
found in FLM89 and new aspherical modes.  As in FLM89, their treatment of
gravity was Newtonian.

In this paper we abandon the Newtonian approximation and adopt a fully
general relativistic treatment of radiation hydrodynamics, motivated by the
known importance of general relativistic effects for the structure of steady
flows.  As in MP95, we adopt a perturbative approach that allows us to find
both spherical and aspherical flow solutions, which are valid as long as the
amplitudes of the flow perturbations are not too large.  Our method, radiation
hydrodynamic equations, and boundary conditions are described in Section~2. 
Because our description of the hydrodynamics is linear, arbitrary disturbances
in the flow can be decomposed into independent modes.  Section~3 contains our
numerical results for the low-order modes.  Our conclusions appear in Section~4.

\section{Description of the Radiation-dominated Flow}

The approach to radiation hydrodynamics adopted here is essentially
perturbative.  It begins with a background flow: a steady, spherically
symmetric solution to the full, nonlinear general relativistic equations for
radiation hydrodynamic accretion by a neutron star.  The radiation hydrodynamic
equations then are linearized about the background flow solution.  The resulting
partial differential equations describe the evolution of small departures from
the background flow.

In the linear regime, an arbitrary initial perturbation can be regarded as a
linear superposition of radiation hydrodynamic eigenmodes.  This enormously
simplifies the investigation of variability and aspherical flow patterns in 
radiation hydrodynamic flows.  To solve for the structure of an eigenmode and its
temporal development, the full system of linearized partial differential
equations is unnecessary; one need only solve an eigenvalue problem based on a
system of ordinary differential equations.

\subsection{General Relativistic Radiation Hydrodynamic Equations}

The flows employed in this study are extremely simple.  Only the gravitation of
the accreting neutron star and radiation forces govern the flow dynamics.  Since
gravitation by the flow itself is insignificant, and the neutron star is assumed
to have negligible rotation, the spacetime outside the stellar radius $R$ is
described by the Schwarzschild metric
\begin{equation}\label{metric}
ds^2=g_{\alpha\beta}dx^\alpha dx^\beta =
\left(1-r_s/r\right)dt^2-\left(1-r_s/r\right)^{-1}dr^2
-r^2\,d\theta^2-r^2\sin^2\theta\,d\phi^2\,,
\end{equation}
where $r$ is the radial coordinate, $\theta$ and $\phi$ are the usual spherical
angular coordinates, and the gravitational radius $r_s$ of the neutron star is
$r_s=2GM/c^2$ in terms of the neutron star mass $M$.  (We adopt units in which
$c\equiv 1$, but for the sake of clarity occasionally retain factors of $c$ in
mathematical expressions.)

Within this spacetime, momentum and energy are conserved locally, so that
\begin{equation}\label{premomentumenergy}
T^{\alpha\beta}{}_{;\beta}=0\,,
\end{equation}
where $T$ is the stress-energy tensor ($T^{\alpha\beta}$ is the flux of the
$\alpha$-momentum component in the $\beta$-direction; see Weinberg 1972) and the
semicolon denotes covariant differentiation.  The stress-energy tensor is the
sum of a part $T_{\rm g}$ due to the gas and a part $T_{\rm ph}$ due to photons. 
Thus,
\begin{equation}\label{momentumenergy}
T_{\rm g}^{\alpha\beta}{}_{;\beta}=-T_{\rm ph}^{\alpha\beta}{}_{;\beta}
\equiv A^\alpha\,.
\end{equation}
The four-vector $\vec A$ describes the exchange of momentum and energy between
the gas and the radiation field.  At the temperatures ($\sim 1$--$10\,{\rm
keV}$) and densities ($\lta 10^{-2}{\rm g\,cm}^{-3}$) typical of the flows
considered in this paper, the radiation energy density grossly exceeds the
thermal energy density of gas particles in the flow.  Moreover, to a good
approximation, the opacity $\kappa$ of the gas is constant and independent of
photon energy, since it is dominated by Thomson scattering, and the number of
free electrons per baryon does not vary (electron-positron pair production is
unimportant, and it is assumed that the gas is fully ionized).  The former
condition allows the thermal energy of the gas to be neglected, so that the
stress-energy tensor for the gas is
\begin{equation}\label{gasstressenergy}
T_{\rm g}=\rho\vec u\otimes\vec u\,,
\end{equation}
where $\vec u$ is the four-velocity of the gas and $\rho$ is the mass density of
the gas in the comoving frame.  Since the thermal energy of the gas is
negligible, Compton heating and cooling terms do not appear in $\vec A$, which
describes only the effects on the gas of the bulk radiation force: if $\vec
a=d\vec u/d\tau$ is the four-acceleration of the gas, then $\vec A=\rho\vec a$. 
In the comoving frame the gas is instantaneously at rest, and as Compton
scattering transfers momentum from the radiation field to the gas, the gas
experiences a three-acceleration $\vec a_{\rm co} = \vec F_{\rm co}\kappa/c$,
where $\vec F_{\rm co}$ is the radiation energy flux (the subscript ``co''
refers to quantities measured by a comoving observer).  This determines $\vec
A$, which is
\begin{equation}\label{aprescription}
\vec A = \rho\vec a
= \left[\vec u\cdot T_{\rm ph}-
\left(\vec u\cdot T_{\rm ph}\cdot\vec u\right)\vec u\right]
{\kappa\over c}\rho\,.
\end{equation}
(The quantity in square brackets is the four-vector generalization of $\vec
F_{\rm co}$.)

The flux of baryons in the gas is $\vec j/m_B$, where $m_b$ is the baryon mass and $\vec j=\rho\vec u$.  When eqn.~[\ref{momentumenergy}] is supplemented by the continuity equation
\begin{equation}\label{continuity}
j^\alpha{}_{;\alpha}=0\,,
\end{equation}
one obtains a complete set of equations for the evolution of the gas flow. 
These equations derive from conservation laws (eqns.~[\ref{momentumenergy}]
and~[\ref{continuity}]), a constitutive relation for the stress-energy tensor
$T_{\rm g}$ in terms of the gas variables $\rho$ and $\vec u$
(eqn.~[\ref{gasstressenergy}]; essentially an ``equation of state''), and a
prescription for the transfer of momentum and energy between the gas and
radiation field (eqn.~[\ref{aprescription}]).

To complete the description of the radiation field, a constitutive relation for
$T_{\rm ph}$ is required.  Close to the stellar surface, where the photon
scattering mean-free paths $l_{\rm mfp}$ typically are short compared to the
length scales $l_{\rm flow}$ characterizing gradients in the radiation energy
density or flow velocity, the radiation field seen by an observer moving with
the gas is almost isotropic.  In these optically thick regions the flow speed
$v$ measured by a stationary observer satisfies $v\ll c$, and to a good
approximation the radiation stress-energy tensor is
\begin{eqnarray}\label{tphdiffusion}
T_{\rm ph}&\approx& E\left(1-r_s/r\right)^{-1}\vec e_t\otimes\vec e_t
\nonumber\\
&&+F_r\left(\vec e_r\otimes\vec e_t+\vec e_t\otimes\vec e_r\right)
+F_\theta r^{-1}\left(1-r_s/r\right)^{-1/2}
\left(\vec e_\theta\otimes\vec e_t+\vec e_t\otimes\vec e_\theta\right)
\nonumber\\
&&+F_\phi (r\sin\theta)^{-1}\left(1-r_s/r\right)^{-1/2}
\left(\vec e_\phi\otimes\vec e_t+\vec e_t\otimes\vec e_\phi\right)\nonumber\\
&&+{E\over 3}
\left(\left[1-r_s/r\right]\vec e_r\otimes\vec e_r+
r^{-2}\vec e_\theta\otimes\vec e_\theta
+[r\sin\theta]^{-2}\vec e_\phi\otimes\vec e_\phi\right)
\,,
\end{eqnarray}
where $E$ is the radiation energy density, and $F_r$, etc., are the components of
the radiation energy flux, all measured by a stationary observer (see
Park~1993 for a discussion of the spherical case).  The basis vectors $\vec
e_\alpha$ are not normalized, since they are coordinate-based: $\vec
e_\alpha\cdot\vec e_\beta=g_{\alpha\beta}$ (see, e.g., Shapiro and
Teukolsky~1983).  Terms omitted from expression~(\ref{tphdiffusion}) are $\lta
(l_{\rm mfp}/l_{\rm flow})(v/c)E$ and $\lta (v/c)^2E$; such terms include the
small corrections to the radiation force on the gas flow due to radiation
viscosity.  Far from the stellar surface the flow is optically thin, radiation
streams radially outward, and the radiation stress tensor is %
\begin{equation}\label{tphstreaming} T_{\rm ph}\approx
E\left(1-r_s/r\right)^{-1}\vec e_t\otimes\vec e_t +F_r\left(\vec e_r\otimes\vec
e_t+\vec e_t\otimes\vec e_r\right) +E\left(1-r_s/r\right)\vec e_r\otimes\vec e_r
\,.
\end{equation}
Expression~(\ref{tphdiffusion}) completes the dynamical description of the
radiation field in the optically thick regime, as does
expression~(\ref{tphstreaming}) for the optically thin regime, since when they
are substituted in the conservation equation~(\ref{momentumenergy}), both yield
equations for the temporal evolution of all the components of $T_{\rm ph}$. 

This simplicity is lost in the intermediate regime, where to find $T_{\rm
ph}$ one must generally solve for the entire photon distribution function (the
specific intensity in all directions at every location in the flow).  A simpler
procedure, which prevents the radiation hydrodynamic calculation from becoming
intractable, is to force closure of the equations for the radiation field by
adopting an expression for $T_{\rm ph}$ that interpolates between the optically
thick and optically thin limits.  The form used for the calculations in this
paper is
\begin{eqnarray}\label{tphansatz}
T_{\rm ph}&\approx& E\left(1-r_s/r\right)^{-1}\vec e_t\otimes\vec e_t
\nonumber\\
&&+F_r\left(\vec e_r\otimes\vec e_t+\vec e_t\otimes\vec e_r\right)
+F_\theta r^{-1}\left(1-r_s/r\right)^{-1/2}
\left(\vec e_\theta\otimes\vec e_t+\vec e_t\otimes\vec e_\theta\right)
\nonumber\\
&&+F_\phi (r\sin\theta)^{-1}\left(1-r_s/r\right)^{-1/2}
\left(\vec e_\phi\otimes\vec e_t+\vec e_t\otimes\vec e_\phi\right)\nonumber\\
&&+f_E E\left(1-r_s/r\right)\vec e_r\otimes\vec e_r
+{1-f_E\over 2}E\left(r^{-2}\vec e_\theta\otimes\vec e_\theta
+(r\sin\theta)^{-2}\vec e_\phi\otimes\vec e_\phi\right)
\,.
\end{eqnarray}
The variable Eddington factor $f_E$ is taken to be a time-independent function
of radius $r$, defined through
\begin{equation}\label{fedef}
f_E\equiv {1+\tau\over1+3\tau}\,,
\end{equation}
where $\tau$ is the scattering optical depth of the unperturbed background flow:
\begin{equation}\label{taudef}
\tau\equiv \tau\left(r_b\right)
+\int_r^{r_b}\kappa \vec e_t\cdot\vec j_0\,\left(1-r_s/r\right)^{-1}dr'\,.
\end{equation}
The subscript ``0'' is used here to denote an unperturbed flow quantity.  The
sensitivity of steady flow calculations to the form of the $f_E$ prescription
is discussed in Park~\& Miller~1991.  To test the dependence of our dynamical
results on the $f_E$ prescription, we occasionally repeated our calculations
with the choice $f_E=(1+\tau[1+\tau])/(1+3\tau[1+\tau])$.  We found no
significant differences in the mode eigenvalues $\omega$ (for a description of
the modes and a definition of $\omega$, see eqn.~[\ref{scalartildedef}] and the
following discussion).

\subsection{Boundary Conditions}

The accreting neutron star is centered within a spherical boundary surface of
radius $r_b$.  The outer boundary conditions describe the entry of gas into
the flow volume, and the escape of radiation.  Gas flows inward though the
boundary surface uniformly and with no angular momentum, so that the gas
four-velocity
\begin{equation}\label{udef}
\vec u\equiv u^t\vec e_t+u^r\vec e_r+u^{\hat\theta}r^{-1}\vec e_\theta
+u^{\hat\phi}(r\sin\theta)^{-1}\vec e_\phi
\end{equation}
(the $\theta$- and $\phi$- components have been normalized for later
convenience) and comoving density $\rho$ there are given by
\begin{equation}\label{outerbcv}
\vec u\left(r_b\right)=u_b^t\vec e_t+u_b^r\vec e_r
\end{equation}
and
\begin{equation}\label{outerbcrho}
\rho\left(r_b\right)=\rho_b\,,
\end{equation}
where $u_b^t$, $u_b^r$, and $\rho_b$ are constants (note that $u_b^r<0$, since
gas flows inward through the boundary).  An observer very far from the
neutron star (where gravitational redshifts are unimportant) sees baryons enter
the flow at a rate $\dot N_B = -4\pi r_b^2 u_b^r\rho_b/m_B$.  The mass accretion
rate is
\begin{equation}\label{mdotdef}
\dot M_r\equiv -4\pi r_b^2 u_b^r\rho_b\,.
\end{equation}
The radius $r_b$ is large enough that the probability of photon scattering from
material near the outer boundary is very small ($r_b\rho_b\kappa/m\ll 1$), and
so the radiation may be assumed to stream radially outward there.  Consequently,
the radiation flux at the outer boundary is related to the energy density by
\begin{equation}\label{outerbcrad}
F_r=cE\,.
\end{equation}

The inner flow boundary is also spherical and just encloses the stellar
surface.  Material flows through the inner boundary onto the stellar surface,
where its kinetic energy is converted to radiation.  This radiation enters the
flow though the inner boundary:
\begin{equation}\label{innerbc}
F_r(R)=F_0+\left(1-r_s/R\right)^{1/2}\vec e_r\cdot\vec j
\left(\left[1-r_s/R\right]^{1/2}u^t-1\right)\,.
\end{equation}
The azimuthal fluxes $F_\theta$ and $F_\phi$ are unconstrained.  The constant
term $F_0$ represents a steady contribution to the flux from the stellar surface
that is independent of the instantaneous state of the impinging accretion flow;
e.g., it can be used to account for radiation produced by steady nuclear burning
beneath the neutron star surface.
If the spherical flows studied here are regarded as approximations to a
quasi-spherical coronal flow component that overlies a thin disk flow (see
FLM89, MP95), $F_0$ may be used to mimic the injection of radiation into the
coronal flow from the innermost portions of the embedded disk.
Physically, $F_0$ is an energy flux measured by an observer at the stellar
surface.  Because a stationary observer at infinity finds redshifts in the
photon energy and arrival rate, the $F_0$ term contributes an amount
\begin{equation}\label{lnoughtdef}
L_0\equiv 4\pi R^2 F_0\left(1-r_s/R\right)
\end{equation}
to the luminosity of the accreting system.

These boundary conditions, which describe how material and radiation enter and
leave the flow volume, completely define a steady spherical background flow, and
also fully determine the temporal development of any initial deviations from the
steady flow.

\subsection{Background Flows}

It is convenient to label background flows with two dimensionless parameters. 
The first describes how close the luminosity of the accretion flow is to the
Eddington limiting luminosity $L_E$, the critical point beyond which steady
spherical flow solutions cease to exist:
\begin{equation}\label{epsilondef}
\epsilon\equiv 1-{\dot M_r\over\dot M_E}-{L_0\over L_E}\,,
\end{equation}
where
\begin{equation}\label{mdotedef}
\dot M_E\equiv{L_E c^{-2}\over 1-\left(1-r_s/R\right)^{1/2}}
\end{equation}
is the Eddington accretion rate.  As $\epsilon$ approaches zero, the system
luminosity approaches the Eddington limit.  The second parameter characterizes
the accretion rate in terms of the Eddington accretion rate:
\begin{equation}\label{mudef}
\mu_r\equiv {\dot M_r\over\dot M_E}\,.
\end{equation}

The near-critical background flows have a simple structure (see Park \&
Miller~1991, and references therein).  Far from the neutron star, material
accelerates as it falls inward.  Radiation forces on the infalling gas prevent
it from achieving full free-fall speeds.  Instead, radiation effectively dilutes
the gravitational field of the star and the speed approaches the modified
free-fall value $u^r_{\rm mff}\equiv-(\epsilon2GM/r)^{1/2}$.  At a transition
radius $r_t\sim(2\mu_r/\epsilon)R$, the flow starts to become optically
thick to scattering, and the radiation energy density mounts more rapidly with
decreasing radius than $r^{-2}$.  The radiation drag on the inflowing gas
increases with the radiation energy density, and the flow speed falls below
the modified free-fall value.  At smaller radii, where the flow is quite
optically thick, the radiation drag is a dominant influence on the flow, and the
flow decelerates as it moves inward, settling onto the stellar surface.  Thus,
the flow structure has three regimes: an optically thin regime far from the
neutron star where the flow accelerates inward, a transitional regime, and an
optically thick settling regime close to the neutron star, where the radiation
energy density is high and radiation drag decelerates the inflow.  Figure~1
shows the velocity profile $u^r(r)$ for a typical near-critical accretion flow.

As an aside, we note that the background flow in Figure~1 does not merge smoothly
into the stellar surface.  Within the surface, gas pressure, particularly electron
degeneracy pressure, dominates radiation pressure, while the opposite is true in
the flow above the stellar surface.  The transition between the two limits occurs
abruptly in a layer of thickness $\lta\epsilon_{\em gas} R^2/(GMm)\ll R$, where
$\epsilon_{\em gas}$ is the average random kinetic energy per gas particle, and
$m$ is the average mass of a gas particle.  (If the flow speed exceeds the random
velocities of gas particles, the gas makes the transition to the stellar surface
through a shock.)  To reproduce the transition layer, it would be necessary to
include gas pressure and an energy equation for the gas in the mathematical
description of the flow, and our numerical code would be forced to follow the
flow either through a shock or through an enormous dynamical range of densities
in a geometrically very thin layer.  Since the detailed structure of the
transition layer does not significantly affect the radiation field above the
surface, it is unimportant to the large-scale dynamics of the flow, and we avoid
considerable numerical difficulties by implicity placing the transition just
below our inner boundary at $R$.

\subsection{Radiation Hydrodynamic Modes}

Because the background flows are time-independent and spherically symmetric, an
arbitrary perturbation can be written as a sum of eigenmodes, each of which
depends exponentially on the time coordinate $t$, and as a spherical harmonic
on the angular coordinates $\theta$ and $\phi$.  Hence, scalar quantities such as
the comoving density $\rho$ are written in the form
\begin{equation}\label{perturbationdef}
\rho=\rho_0+\rho_1\,,
\end{equation}
where the subscript~``0'' refers to the background flow value, and the
subscript~``1'' denotes the perturbation.  For an eigenmode,
\begin{equation}\label{scalartildedef}
\rho_1=\tilde\rho(r)\,Y_{lm}(\theta,\phi)\,\exp(-i\omega t)\,.
\end{equation}
The eigenfrequency $\omega$ is complex, and the modes can grow or decay with
time.  Modes with ${\rm Im}(\omega)<0$ are stable, and modes with ${\rm
Im}(\omega)>0$ are unstable.  The radial and time components of the four-velocity
perturbation $\vec u_1$ may also be written in the form~(\ref{scalartildedef}),
as may the radiation energy density perturbation
$E_1$ and radial flux perturbation $F_{r1}$. Quantities such as
$u^{\hat\theta}_1$, $u^{\hat\phi}_1$, $F_{\theta 1}$, and $F_{\phi 1}$ are also
related to the spherical harmonics and take the forms
\begin{equation} \label{thetatildedef}
u^{\hat\theta}_1 =
\utheta(r)\,\partial_\theta Y_{lm}(\theta,\phi)
\,\exp(-i\omega t)
\end{equation}
and
\begin{equation} \label{phitildedef}
u^{\hat\phi}_1 =
\uphi(r)\,{1\over\sin\theta}\partial_\phi Y_{lm}(\theta,\phi)
\,\exp(-i\omega t)\,.
\end{equation}
The spherical symmetry of the background flow implies that the eigenfrequencies
and radial eigenfunctions (e.g., $\den[r]$) depend only on the harmonic
number $l$, not on the azimuthal number $m$.  In the discussion below, we
therefore label modes only by $l$ and suppress $m$.  By decomposing perturbations
into eigenmodes, the problem of finding the time-evolution of weakly unsteady
and possibly aspherical flows is reduced to that of finding the modes, which in
turn involves only solving a system of ordinary differential equations (see
Appendix~A).

The numerical method we used to find the modes in this paper is slightly
different from the one described in MP95.  First, we calculate a background flow
with a given luminosity and mass accretion rate (parameterized by $\epsilon$
and $\mu_r$; as an outer boundary condition, all the background flows in this
paper have velocities $u^r_b=u^r_{\rm mff}[r_b]$).  Next, for a given value of
$l$, we obtain approximate values of the eigenfrequencies for all the
low-frequency modes by making a map of the complex plane.  Each point on the
map corresponds to a {\em trial} eigenvalue.  Using the trial eigenvalue and
the outer boundary conditions, we integrate the linearized radiation
hydrodynamic equations inward and obtain a figure of merit (a real number
$\zeta\ge0$) that quantifies the failure of the trial eigenvalue to satisfy the
inner boundary condition.  Trial values that are true eigenvalues produce
$\zeta(\omega)=0$, and other trials produce higher numbers.  We plot contours
of $\zeta$ on the trial $\omega$-plane (an example appears in Figure~2; only
the right half of the complex plane is plotted, since the equations of motion
guarantee that each eigenfrequency $\omega$ on the right half of the complex
plane has a twin on the left side, at $-\omega^*$), and find the eigenvalues by
visual inspection.  With these visual estimates of the eigenvalues as starting
values, we are able to solve numerically for the modes and their eigenvalues by
simple automated shooting, refining the eigenvalue approximation at each
successive iteration.  This scheme guarantees that no low-frequency eigenmodes
are overlooked.

\section{The Modes and their Frequencies}

\subsection{Numerical Results and Interpretation}

Low-frequency radiation hydrodynamic eigenmodes may be categorized as
oscillatory or nonoscillatory, and as spherical or aspherical.

Oscillatory modes occur through a feedback mechanism in which enhanced radiation
production hinders the progress of material from the outer boundary to the
stellar surface, so that episodes of high radiation production alternate with
intervals during which the inner flow is starved, and radiation production
falls below the average.  The periods of these oscillations are typically
comparable to the inflow time $t_0$, the time taken by a fluid particle in the
unperturbed flow to pass from the outer boundary to the stellar surface, as
measured by an observer at infinity:
\begin{equation} \label{tnoughtdef}
t_0\equiv -\int_R^{r_b}{u^t\over u^r}\,dr\,.
\end{equation}
Oscillatory modes are both spherical ($l=0$) and aspherical ($l>0$).

As an example of an aspherical mode, Figure~3 shows the first half of an $l=2$,
$m=0$ oscillation.  The radial (${\rm Re}[\ur\exp(-i\omega t)]$) and tangential
(${\rm Re}[\utheta\exp(-i\omega t)]$) velocity perturbations at the beginning,
middle, and end of the half cycle are plotted.  Only the radial dependences of
the perturbed velocity components are shown; the components depend on the
colatitude $\theta$ through equations~(\ref{scalartildedef})
and~(\ref{thetatildedef}).  Thus, the full spatial dependence of the perturbed
velocity is given by
\begin{equation}
u^r_1=
{\sqrt5\over2}\,(3\cos^2\theta-1)\,{\rm Re}(\ur\exp[-i\omega t])
\end{equation}
and
\begin{equation}
u^{\hat\theta}_1=
-3\sqrt5\,\cos\theta\sin\theta\,{\rm Re}(\ur\exp[-i\omega t])\,.
\end{equation}
The mode in Fig.~3 has been normalized so that the radiative flux perturbation
at the outer boundary is $F_{r1}(r_b)=(\sqrt5/2)\,(3\cos^2\theta-1)\,{\rm
Re}(\exp-i\omega t)\,L_E/(4\pi r_b^2)$.  The flux is high at $t=0$, the onset
of the half cycle.  At the same instant, a negative perturbation in the radial
velocity increases the mass accretion rate above the stellar poles, and pours
energy into the radiation field.  Eddy currents in the gas advect the excess
radiation out of the flow. The rising radiation pressure drives material
sideways, away from the poles.  Gas moves from both poles across the stellar
surface, until it reaches the equator and is forced upward.  At about $0.3\,R$
above the stellar surface (a distance comparable to the density and pressure
scale heights in the unperturbed flow), the radial velocity perturbations
vanish, and gas which has risen from the stellar equator stalls and drifts
sideways, to join the columns descending onto the poles.  The perturbed flow
pattern close to the stellar surface thus essentially consists of two
doughnut-shaped convective cells, which draw gas down over the stellar poles
and loft gas and radiation above the equator.  Immediately above each of these
two cells sits an elongated counterrotating cell that extends outward into the
optically thin portions of the flow.  This second tier of cells carries
radiation-laden material horizontally back from the equator to the poles, where
it rises.  Thus, above each pole where a descending perturbation enhances the
production of radiation, rising gas in the second tier of convective cells
carries radiation out of the optically thick inner flow, and contributes to the
positive perturbation in radiative flux at the flow's outer boundary.  The
excess radiation flux decelerates gas entering the flow through the outer
boundary, reinforcing the flow perturbation and ensuring that as each
convective cell is carried out of the flow through the inner boundary, a new
cell forms in the gas entering the flow.  Because the feedback cycle is
inefficient, each repetition is weaker than the one that preceded it, and the
mode decays.  The velocity perturbations at the middle and the end of the half
cycle show how the pattern of alternating cells weakens as the background flow
advects it inward.

Spherical nonoscillatory modes are related to the spherical oscillatory modes
and typically are damped.  An analytic calculation for the frequencies of
spherical modes in optically thin flows gives
\begin{equation} \label{frequencyestimate}
{\omega t_0\over 2\pi}\approx
n-i\ln\left(1+{5\over 2}{\epsilon\over\mu_r}\right)\,,
\end{equation}
where $n$ is an integer (MP95).  The nonoscillatory modes correspond to $n=0$. 
The flow in a spherical nonoscillatory mode moves inward more slowly than the
unperturbed flow, and is overdense and overluminous.  The luminosity
perturbation inhibits the inward flow of gas, causing a ``traffic jam.''  As
overdense material is accreted by the neutron star, it is replaced by somewhat
less dense material from the outer boundary, and the flow congestion is
gradually relieved.  As in the case of the oscillatory modes, the timescale for
the evolution of the mode is set by the flow time $t_0$.

The aspherical nonoscillatory modes are often unstable, and correspond to the
onset of global convective motions in the optically thick inner flow.  Figure~4
shows the velocity perturbations of an unstable $l=2$, $m=0$ mode at three
successive moments in its development.  (The normalization of the mode is the
same as in Fig.~3.)  The velocity perturbations are similar to those seen at the
beginning of the oscillation in Fig.~3, but the pattern in a nonoscillatory mode
rises upward through the descending background flow just fast enough to remain
fixed in space.  Rather than being advected by the flow, it simply grows in
strength.

The growth rates of the aspherical nonoscillatory modes are determined in the
inner parts of the accretion flow, and are consequently independent of the
flow time $t_0$ (which is dominated by the time a fluid particle spends
traveling through the outer parts of the flow) for sufficiently large outer
boundary radii $r_b$.  Tables 1--6 show how, as the outer boundary radius
increases from $10^2R$ to $10^3R$, the growth rates of the aspherical
nonoscillatory modes approach finite asymptotic limits.  This contrasts strongly
with the behavior of the oscillatory and spherical nonoscillatory modes
frequencies, which decrease to zero as $r_b\rightarrow\infty$.  {\em Thus,
independent of the boundary conditions imposed at large distances from the
accreting star, spherical radiation hydrodynamic accretion is unstable to
aspherical perturbations --- even at luminosities significantly below the
Eddington limit, where steady, spherically symmetric flow solutions are readily
found.}

The growth of aspherical perturbations appears to be related to the entropy
stratification of the optically thick flow.  Where less buoyant, entropy-poor
material overlies more buoyant, entropy-rich material, the flow is unstable to
convection.  If radiative diffusion is not too severe, the condition for
convective stability is the familiar Schwarzschild criterion, $\Gamma\equiv
d\ln p/d\ln\rho=(d\ln p/dr)/(d\ln\rho/dr)<4/3$.  The background flows
characteristically have a layer in which $\Gamma>4/3$ at moderate optical
depths, near the transition radius $r_t$.  As aspherical perturbations are
carried by the background flow into this layer, they grow.  They then enter the
underlying stably stratified fluid, and execute damped, buoyancy-supported
oscillations --- as seen by an observer moving with the fluid.  An observer at
rest sees a standing wave.  The standing wave grows because it produces
radiation flux perturbations that excite larger ``seed'' velocity perturbations
in the optically thin outer flow.  These subsequently grow in the destabilizing
layer, and ultimately produce yet greater flux perturbations.  The value of the
stability parameter $4/3-\Gamma$ and of the proper radial velocity perturbation
it produces are shown in Figure~5.

(Among all of the general relativistic flows we have examined, we have found no
unstable oscillatory modes, although in MP95 we reported a few weakly unstable
aspherical oscillatory Newtonian modes in flows with luminosities within 2\% of
the Eddington limit [$\epsilon=0.02$].  We have traced the oscillatory
instability to an error in the Newtonian mode-generating program used for that
paper, which significantly affected the eigenfrequencies of only the aspherical
modes of the $\epsilon=0.02$ flows.)

The eigenfrequencies of the modes follow two simple trends: (i)~slower, denser
flows (flows with smaller values of $\epsilon$ or higher values of $\mu_r$)
tend to oscillate more slowly than faster, more tenuous flows; and
(ii)~the radiation hydrodynamic modes in slower, denser flows tend to be less
damped (or even more unstable).  Although they are true of all modes, both
trends are occasionally violated by aspherical modes.  For example, the
frequency of the oscillatory $l=1$ mode in Table~7 {\em rises} slightly as the
system luminosity climbs from $0.6\,L_E$ ($\epsilon=0.4$) to $0.7\,L_E$
($\epsilon=0.3$), although the the higher luminosity flow moves more slowly.

The anomalous behavior of the aspherical modes arises because aspherical waves
propagating through optically thick gas do not have to move at the same speed
as the gas in the flow.  An ingoing aspherical wave thus can take considerably
longer than the flow time $t_0$ to travel from the outer boundary to the stellar
surface, and $t_0$ becomes a poorer estimate of the fundamental timescale for
feedback-supported oscillations.  General disturbances in the
radiation-dominated gas may be decomposed into sound waves (vorticity-free
velocity perturbations, and adiabatic perturbations in the gas and radiation
energy densities) and internal waves (perturbations in the vorticity and entropy
of the flow).  Sound waves travel very rapidly through the optically thick
flow, which is subsonic, and are strongly damped by radiative diffusion in the
optically thin flow.  Sound-like modes consequently are expected to have high
frequencies and to be strongly damped; they play little role in the
low-frequency spherical and aspherical modes, which are essentially internal
waves.  To see how internal waves propagate, consider a wave that varies in
space and time as $\exp\,i(\vec k\cdot\vec x-\omega t)$.  For simplicity,
assume that the wavelength is small compared to typical flow length scales
(such as the density scale height $h=g_{rr}^{-1/2}d\ln\rho/dr$; $kh\ll 1$), and
that radiation diffusion can be neglected.  The frequency $\omega$ and
wavevector $\vec k$ are related through the dispersion relation
\begin{equation} \label{internaldispersion}
\omega = \vec v_0\cdot\vec k+\omega_{\rm BV}
{k_\perp\over k}\,,
\end{equation}
where $\vec v_0$ is the velocity of the unperturbed flow, $\omega_{\rm
BV}\approx([4/3]-\Gamma)^{1/2}(g/h)^{1/2}$ is the Brunt-V\"ais\"al\"a frequency,
and $k_\perp$ is the component of the wavevector $\vec k$ perpendicular to $\vec
g$, the gravitational acceleration.  (The Brunt-V\"ais\"al\"a frequency
describes buoyancy-driven oscillations in a fluid with a stable entropy
stratification.  When the stratification is unstable --- more buoyant fluid
underlying less buoyant fluid --- the frequency is imaginary, and describes the
growth of convective motions.  For a general treatment of internal
waves, see Lighthill [1978]; internal waves in a diffusive, radiation-dominated
fluid are discussed in Miller \& Grossman [1997].  Diffusion tends to reduce the
Brunt-V\"ais\"al\"a frequency and damps the oscillations.)  The important feature
of the dispersion relation is that globally spherically symmetric disturbances
correspond to $k_\perp=0$, and move passively with the fluid.  Aspherical waves
($k_\perp\ne0$) move in the fluid's rest frame, and if they move upstream can
spend a much longer time in the optically thick accretion flow than can spherical
waves, causing the eigenfrequency of the mode to depart significantly from the
estimate~(\ref{frequencyestimate}).  Such anomalous aspherical waves also are
particularly susceptible to general relativistic effects, which are strongest
near the stellar surface.

The eigenfrequencies of general relativistic and Newtonian modes are compared in
Table~7.  The spherical modes are affected only weakly by general relativity. 
This reflects the fact that the key to their operation is the influence of
escaping radiation on the inflowing gas near the outer boundary, where
specifically general relativistic effects are of minimal importance: the mode
eigenfrequencies of the general relativistic and Newtonian flows do not differ
significantly.

In contrast, general relativity often exerts a considerable influence on the
frequencies of the aspherical modes.  As an extreme example, it sometimes
happens that, as the parameter $\epsilon$ is changed, a complex eigenfrequency
$\omega$ on the right half of the complex plane and its twin $-\omega^*$ on
the left half abruptly move to the imaginary axis and merge, and then bifurcate
into two imaginary (nonoscillatory) eigenfrequencies.  This exchange of
oscillatory for nonoscillatory modes occurs in both general relativistic and in
Newtonian flows, but tends to happen at very different values of $\epsilon$ in
the two cases.  Thus, a nonoscillatory general relativistic mode will sometimes
have no Newtonian counterpart.  An example occurs in the $l=1$, $\epsilon=0.3$
entry of Table~7, where a Newtonian 5.2~Hz oscillation and its complex conjugate
have replaced the the pair of nonoscillatory modes present at smaller values of
$\epsilon$.

The rapid drop of the stability parameter with radius in Fig.~5 raises the
possibility that a class of aspherical oscillations exists that we have not yet
found in our numerical surveys.  Along with the stability parameter, the
Brunt-V\"ais\"al\"a frequency decreases steeply with increasing radius,
a condition that should trap high-frequency internal waves near the stellar
surface and give rise to a set of aspherical modes (a terrestrial example
occurs when internal modes are trapped by oceanic thermoclines; Lighthill
1978, p.~302).  We plan to return to this point in future work.

\subsection{Application to Radiation Hydrodynamic QPO Models}

The aspherical radiation hydrodynamic modes share the general tendency of the
spherical modes to decrease in frequency when the system luminosity increases,
but there are anomalous luminosity intervals in which mode frequencies rise
with the luminosity.  An additional peculiarity of the aspherical modes is
that a pair of nonoscillatory modes can coalesce in frequency, and reemerge as
a pair of oscillatory modes.  When this happens, the oscillatory mode frequency
changes very rapidly with slight changes in the luminosity.  Although similar
behavior has been observed in the ``normal branch'' quasiperiodic oscillations
(NBO) of some accreting neutron star systems, and we are partisan supporters of
radiation hydrodynamic models for these models (Lamb 1989, FLM89), we suspect
that aspherical modes are not directly involved.  It appears that as the mass
accretion rate increases by $\gta\,10$\%, the NBO frequency remains almost
constant at $\nu_{\rm QPO}\sim 6\,$Hz, and then increases sharply to
$\gta\,20\,$Hz (van der Klis 1996, pp.~279--285).  Purely spherical radiation
hydrodynamic modes can account for the frequency scales and spectral appearance
(Miller \& Lamb 1992) of the oscillations, but are challenged by the observed
frequency variations.  An aspherical mode could increase that rapidly in
frequency, but would do so starting from $\nu_{\rm QPO}=0$.  It is more likely
that, if the radiation hydrodynamic model is correct, the NBO is a quasispherical
mode in a mildly aspherical coronal flow.  The asphericity of the background flow
would couple spherical and aspherical modes, and (especially given the
instability described above) the coupling could be expected to become more severe
as the system approached the Eddington luminosity.  If a spherical mode with
unperturbed frequency $\nu_s$ couples dominantly to an aspherical mode with
frequency $\nu_a=0+i\alpha$, then to lowest order in the perturbation, the
frequency of the resulting quasispherical mode is $\nu_{\rm
pert}\approx\nu_s+\eta/(\nu_s-i\alpha)$, where $\eta$ describes the strength of
the aspherical coupling.  If the coupling strengthens sufficiently as the
luminosity rises, the quasispherical mode can reproduce the observed frequency
behavior.  Observationally, the steep rise in QPO frequency coincides with a
sudden change in the relation between the mass accretion rate of the source and
its X-ray spectrum.  We argue below that the change in the spectral behavior may
be due to the onset of an aspherical mode instability.

\section{Conclusions}

Our numerical results indicate that spherical, radiation-hydrodynamic accretion
is unstable to aspherical perturbations even when the luminosity $L_\infty$ seen
by an observer at infinity is substantially less than the Eddington limit ---
flows with luminosities as small as $0.6\,L_E$ are mildly unstable.  Within the
limits of a linear stability analysis, it is impossible to say whether the
nonlinear development of the instability culminates in gentle convective
motions within an almost spherical inflow, or whether the flow eventually
differentiates itself into accreting and outflowing zones, e.g., accreting
through the equatorial plane, and outflowing at the poles.
The eigenmodes themselves suggest that the former possibility is the most
likely: convective motions comparable to the unperturbed flow speed might be
expected to ``mix away'' the destabilizing layer on which the mode's growth
depends, but in the optically thin outer flow the perturbations in velocity
still would be well below the mean inflow speeds, and the flux perturbations
would be extremely minor.  Nevertheless, the possibility of radiation-driven
outflow cannot be excluded.  The question can only be resolved with a fully
nonlinear study of two- or three-dimensional radiation hydrodynamic flows.

Whatever the endstates of the unstable radiation hydrodynamic modes might be,
we note speculatively that they may already have been observed in the brightest
low-mass X-ray binaries.  As an accreting system approaches the Eddington
luminosity from below, the growing radiation forces cause the flow to
decelerate and become more optically thick to Compton scattering.  If the
unstable aspherical radiation hydrodynamic modes reach sufficient amplitudes,
this trend may reverse beyond a threshold luminosity: gas motions associated
with the mode advect photons out of the flow more efficiently than would be
possible through diffusion alone, and the average number of scatters suffered
by escaping photons decreases.  Interestingly, detailed models of X-ray
spectral production in the coronae of rapidly accreting low-mass X-ray binaries
{\em require} that when the luminosities $L_\infty$ in these systems closely
approach or exceed $L_E$, the effective scattering optical depth of their coronae
must decrease (Psaltis, Lamb, \& Miller 1995).  Without an instability, at high
accretion rates the coronal flows would simply become progressively glutted with
plasma, and it would be hard to understand the observed spectral behavior.

\acknowledgements

This work was supported in part by NASA grants NAGW-2935 and NAG5-3396, and by
KOSEF grant 971-0203-013-2.

\vfil\eject

\appendix
\setcounter{equation}{0}
\section{Linearized Flow Equations} \label{app:lineq}

Gas enters the outer boundary steadily and uniformly, so that the velocity and
density perturbations vanish there:
\begin{equation}\label{ubounds}
\ur\left(r_b\right)=\utheta\left(r_b\right)=0
\end{equation}
and
\begin{equation}\label{rhobounds}
\den\left(r_b\right)=0
\,.
\end{equation}
Because radiation is streaming outward at the outer boundary,
\begin{equation}\label{froutbounds}
\fr\left(r_b\right)=\energy\left(r_b\right)
\,.
\end{equation}
At the inner boundary, radial flux perturbations reflect perturbations in the
kinetic energy deposited by the flow:
\begin{equation}\label{frinbounds}
\fr(R) =
\left(\left[1-r_s/R\right]^{-1/2}-u^t_0\right)
\left(\den u_0+\ur\rho_0\right)
-u^t_1 u_0\rho_0
\,,
\end{equation}
where the abbreviation $u_0\equiv u^r_0$ has been introduced, and
$u^t_1=\left(1-r_s/r\right)^{-2}\left(u_0/u^t_0\right)\ur$.  These boundary
conditions and the modal angular dependences (eqns.~[\ref{thetatildedef}]
and~[\ref{phitildedef}]) ensure that $\uphi=\utheta$ and $\fphi=\ftheta$; hence
only equations for $\utheta$ and $\ftheta$ appear below.

The radial variations of the perturbed flow quantities are determined by the
boundary conditions above and the following linearized flow equations:
\begin{equation} \label{urderiv}
{d\,\over dr} \ur = u_0^{-1}\left(i\omega \utzero\ur -
u_0'\ur + \kappa\fcor\right)
\,,
\end{equation}
\begin{equation} \label{uthetaderiv}
{d\,\over dr} \utheta = u_0^{-1}\left(i\omega\utzero\utheta -
{u_0\over r}\utheta + \kappa\fthetaco\right)
\,,
\end{equation}
\begin{eqnarray} \label{rhoderiv}
{d\,\over dr} \den &=& u_0^{-1}
\left(i\omega\utzero-u_0'-{2u_0\over r}\right)\den
+ i\omega{\rho_0\over u_0}\ut\nonumber\\&&
+ {u_0'\over u_0^2}\rho_0\ur  - {\rho_0\over u_0}{d\,\over dr}\ur
+ {l(l+1)\over r}{\rho_0\over u_0}\utheta
\,,
\end{eqnarray}
\begin{equation} \label{ederiv}
{d\,\over dr} \energy =
f_E^{-1}\left[
\begin{array}{l}
\left(\left[1-3f_E\right]r^{-1}-f_E'
-\left[1+f_E\right]r_s\left[2r^2\right]^{-1}\left[1-r_s/r\right]^{-1}\right)
\energy\\
-\kappa
\left(u^t_0 F_{\rm co 0r}\den+\rho_0\fcor\right) +i\omega\grr\fr
\end{array}
\right]
\,,
\end{equation}
and
\begin{eqnarray} \label{fderiv}
{d\,\over dr} \fr &=& -\left([2/r]
+\left[r_s/r^2\right]\left[1-r_s/r\right]^{-1}\right)\fr
+i\omega\grr\energy +l(l+1)r^{-1}\ftheta\nonumber\\
&&-\kappa\grr
\left(u_0F_{\rm co 0r}\den+\rho_0F_{\rm co 0r}\ur+\rho_0u_0\fcor\right)
\,;
\end{eqnarray}
where
\begin{equation} \label{utdef}
\ut = \left(1-r_s/r\right)^{-2} {u_0\over u^t_0}\,\ur
\,,
\end{equation}
\begin{equation} \label{fcordef}
\fcor = \grr\left(F_{\rm co 0r}\,\ut + u^t_0\,\Fcor\right)
\,,
\end{equation}
\begin{equation} \label{Fcordef}
\Fcor = \left(
\begin{array}{l}
\left[1+2\grr u_0^2\right]\fr - u_0u^t_0\left(1+f_E\right)c\energy\\
-\left(1+f_E\right)E_0\left[u^t_0\ur+u_0\ut\right]\\
+\left[4\grr u_0 F_{\rm 0r} - u^t_0\left(1+f_E\right)E_0\right]\ur
\end{array}
\right)
\,,
\end{equation}
\begin{equation} \label{fthetacodef}
\fthetaco = \grr u^t_0\ftheta
-\left(E_{\rm co 0}+\left[1-f_E/2\right]E_0\right)
\utheta
\,,
\end{equation}
\begin{equation} \label{fthetadef}
	\ftheta = \left({u^t_0\over\left[1-r_s/r\right]}
-{i\omega\over\kappa\rho_0}\right)^{-1}
\left[{f_E-1\over 2r\kappa\rho_0 u^t_0}\energy
+\left(E_{\rm co 0}+{1-f_E\over 2}E_0\right)\utheta\right]
\,,
\end{equation}
\begin{equation} \label{ecodef}
E_{\rm co 0} = E_0-2F_{\rm r 0}\utzero u_0
+ \grr\left(1+f_E\right)u_0^2 E_0
\,,
\end{equation}
and derivatives of background quantities are primed ($u_0'\equiv du^r_0/dr$ and
$f_E'\equiv df_E/dr$).

\newpage

\def\aa{A\&A}
\def\apj{ApJ}
\def\araa{ARA\&A}
\def\mnras{MNRAS}
\def\nat{Nature}
\def\pasj{PASJ}
\def\ssr{Space Sci.~Rev.}

\clearpage

\begin{figure}
\caption
{
The velocity profile $u^r(r)$ of a typical near-critical accretion flow
($\epsilon=0.2$, $\mu_r=0.8$).  At large radii, the velocity profile is
approximately that of freely falling material in a dilute gravitational field:
$u^r(r)\approx u_{\rm mff}(r)\equiv(\epsilon r_s/r)^{1/2}=(\epsilon
2GM/[c^2r])^{1/2}$.  The velocity profile flattens at intermediate radii
$r\sim(2\mu_r/\epsilon)R$, where the flow becomes optically thick to Thomson
scattering.  The flow decelerates at smaller radii, and ``settles'' onto the
stellar surface.
}
\end{figure}

\begin{figure}
\caption{
This contour plot shows the location of low-frequency $l=2$ mode frequencies in
the complex plane.  Only the half of the plane with $\Re(\omega)>0$ is shown,
since the contours are symmetrical under reflections through the imaginary
axis.  Mode frequencies appear as the deepest points in valleys on the contour
map, where the figure of merit $\zeta=0$.  The background flow accretes material
rapidly ($\mu_r=0.8$) and has a luminosity near the Eddington limit
($\epsilon=0.2$).  The outer flow boundary is at $r_b=100R$.  The unit of
frequency here is $2\pi/t_0$, where $t_0$ is the inflow time, the length of time
taken for a fluid particle in the unperturbed flow to travel from the outer
boundary to the stellar surface.  A mode that repeats once per inflow time has
$\Re(\omega)=1$.  Note the presence of an eigenmode at $\omega=0.000+0.148i$,
which shows the flow is globally unstable.
}
\end{figure}

\begin{figure}
\caption
{
The structure of a typical aspherical ($l=2$, $m=0$) oscillatory mode, in a
background flow close to the Eddington limit ($\epsilon=0.2$, $\mu_r=0.8$).
Radial and tangential velocity perturbations are shown at the beginning (solid
line, and the line interrupted by single short dashes), middle (dotted and
double-dashed lines), and end (dashed and triple-dashed lines) of the first half
of the oscillation cycle.  Only the radial dependences of the velocity
perturbation components are plotted; the full three-dimensional velocity patterns
are given by $(\sqrt5/2)(3\cos^2\theta-1)$ times the plotted radial perturbations
(solid, dotted, and dashed lines), and by $-3\sqrt5\sin\theta\cos\theta$ times
the plotted tangential perturbations (lines interrupted by single, double, and
triple-dashes).  The velocity perturbations at the beginning of the half cycle
take the form of convective cells near the stellar surface.  These cells enhance
the flow of material onto the stellar poles, and reduce the accretion rate just
above the equator.  Near the stellar surface, radiation flows outward
preferentially through the stalled material over the equator.  A second tier of
convective cells overlies the first, and diverts the flow of radiation back
toward the polar axis of the flow.  Radiation escaping along the polar axis
inhibits the inward flow of gas from the poles of the outer boundary surface, and
encourages accretion from the equator.  Thus, as the background flow advects one
tier of convective perturbations out of the flow volume through the inner
boundary, perturbations in the radiation flux seed the formation of a new tier of
cells in the flow close to the outer boundary, and a cycle of alternatingly
enhanced and diminished accretion is established.  Because the feedback mechanism
that supports the cycle is inefficient, the perturbations decay with time.
}
\end{figure}

\begin{figure}
\caption
{
The structure of an unstable aspherical ($l=2$, $m=0$) nonoscillatory mode, in a
background flow close to the Eddington limit ($\epsilon=0.2$, $\mu_r=0.8$).
Radial and tangential velocity perturbations are shown at three stages in the
development of the growing mode.  Although the velocity perturbations resemble
those of the oscillatory mode in Fig.~3, the pattern of velocity perturbations
is not dragged into the accreting neutron star with the background flow; it
simply grows with time.
}
\end{figure}

\begin{figure}
\caption
{
The growth mechanism of an unstable aspherical ($l=2$, $m=0$) nonoscillatory
mode.  The background flow has a luminosity at infinity $L_\infty=0.9L_E$
(dimensionless parameters $\epsilon=0.1$, $\mu_r=0.9$), closer to the Eddington
limit than the flows of Figures~1--4.  The radial velocity perturbation appears
as a solid line.  As in Fig.~3, only the radial dependence of the velocity
perturbation component is plotted.  The convective stability parameter
$4/3-\Gamma$ is plotted as a dashed line.  Aspherical perturbations carried
inward by the accretion flow grow in the layer $7R<r<25R$, where the stability
parameter is negative.  At radii $r<7R$ the flow is stable against convective
motions; a fluid element adiabatically displaced downward rises again because
it is less dense than its surroundings.  An observer moving with the background
flow sees perturbed gas elements bob up and down.  An observer at rest sees the
standing wave pattern in the Figure.  The standing wave grows because the wave
drives radiation flux perturbations in the optically thin outer flow, which in
turn become the ``seeds'' that will become amplified in the flow's
destabilizing layer, and feed the standing wave.
}
\end{figure}

\clearpage

\begin{deluxetable}{rrcc}
\footnotesize
\tablecaption{Eigenfrequencies: $l=0$ Modes}
\tablewidth{30pc}
\tablehead{
          \colhead{$\epsilon$}  & \colhead{$\mu_r$}     
         & \colhead{$\nu t_0$}  & \colhead{$\nu$ (Hz)}
}
\startdata
0.10 & 0.90 & 0.0000 -0.0303i & 0.000 -0.212i \nl
     &      & 0.8260 -0.2284i & 5.768 -1.595i \nl
     &      & 1.8697 -0.2758i & 13.056 -1.926i \nl
0.20 & 0.80 & 0.0000 -0.0933i & 0.000 -1.076i \nl
     &      & 0.8760 -0.1493i & 10.106 -1.722i \nl
     &      & 1.8051 -0.1729i & 20.824 -1.995i \nl
0.30 & 0.70 & 0.0000 -0.1492i & 0.000 -2.195i \nl
     &      & 0.9102 -0.1732i & 13.392 -2.548i \nl
     &      & 1.8365 -0.1823i & 27.021 -2.683i \nl
0.40 & 0.60 & 0.0000 -0.2000i & 0.000 -3.466i \nl
     &      & 0.9200 -0.2129i & 15.946 -3.691i \nl
     &      & 1.8466 -0.2183i & 32.006 -3.783i \nl
0.30 & 0.30 & 0.0000 -0.2474i & 0.000 -3.655i \nl
     &      & 0.9342 -0.2630i & 13.802 -3.886i \nl
     &      & 1.8749 -0.2675i & 27.699 -3.951i \nl
0.20 & 0.30 & 0.0000 -0.1907i & 0.000 -2.224i \nl
     &      & 0.9203 -0.2236i & 10.733 -2.607i \nl
     &      & 1.8618 -0.2362i & 21.715 -2.755i \nl
0.10 & 0.30 & 0.0000 -0.0708i & 0.000 -0.479i \nl
     &      & 0.8451 -0.3778i & 5.723 -2.559i \nl
     &      & 1.9938 -0.4725i & 13.503 -3.200i \nl
\enddata
\tablecomments{Boundary conditions: $r_b=100R$, 
               $u_b=u_{\rm mff}(r_b)$}
\end{deluxetable}

\begin{deluxetable}{rrcc}
\footnotesize
\tablecaption{Eigenfrequencies: $l=1$ Modes}
\tablewidth{30pc}
\tablehead{
          \colhead{$\epsilon$}  & \colhead{$\mu_r$}     
         & \colhead{$\nu t_0$}  & \colhead{$\nu$ (Hz)}
}
\startdata
0.10 & 0.90 & 0.4545 -0.6832i & 3.174 -4.771i \nl
     &      & 2.0815 -0.6168i & 14.535 -4.308i \nl
     &      & 3.2723 -0.6420i & 22.851 -4.483i \nl
0.20 & 0.80 & 0.0000 -0.1107i & 0.000 -1.277i \nl
     &      & 0.5889 -0.8865i & 6.794 -10.226i \nl
     &      & 1.6473 -0.5056i & 19.003 -5.833i \nl
     &      & 2.6987 -0.4823i & 31.132 -5.564i \nl
0.30 & 0.70 & 0.0000 -0.1343i & 0.000 -1.976i \nl
     &      & 0.4839 -0.6444i & 7.120 -9.481i \nl
     &      & 1.5573 -0.4251i & 22.913 -6.254i \nl
     &      & 2.5671 -0.4072i & 37.771 -5.991i \nl
0.40 & 0.60 & 0.0000 -0.3028i & 0.000 -5.249i \nl
     &      & 0.5035 -0.4871i & 8.727 -8.442i \nl
     &      & 1.5491 -0.4040i & 26.849 -7.003i \nl
     &      & 2.5307 -0.3904i & 43.862 -6.766i \nl
0.30 & 0.30 & 0.0000 -0.1445i & 0.000 -2.135i \nl
     &      & 0.5046 -0.4838i & 7.455 -7.147i \nl
     &      & 1.5528 -0.3907i & 22.941 -5.772i \nl
     &      & 2.5482 -0.3760i & 37.646 -5.554i \nl
0.20 & 0.30 & 0.0000 -0.0274i & 0.000 -0.320i \nl
     &      & 0.5574 -0.6083i & 6.501 -7.094i \nl
     &      & 1.6134 -0.4324i & 18.817 -5.043i \nl
     &      & 2.6435 -0.4095i & 30.831 -4.776i \nl
0.10 & 0.30 & 0.0000 -0.2877i & 0.000 -1.949i \nl
     &      & 0.7767 -0.8089i & 5.260 -5.479i \nl
     &      & 2.0833 -0.5440i & 14.109 -3.684i \nl
\enddata
\tablecomments{Boundary conditions: $r_b=100R$, 
               $u_b=u_{\rm mff}(r_b)$}
\end{deluxetable}

\begin{deluxetable}{rrcc}
\footnotesize
\tablecaption{Eigenfrequencies: $l=2$ Modes}
\tablewidth{30pc}
\tablehead{
          \colhead{$\epsilon$}  & \colhead{$\mu_r$}     
         & \colhead{$\nu t_0$}  & \colhead{$\nu$ (Hz)}
}
\startdata
0.10 & 0.90 & 0.0000 +0.0018i & 0.000 +0.013i \nl
     &      & 1.1938 -0.5557i & 8.336 -3.880i \nl
     &      & 2.4761 -0.4908i & 17.291 -3.428i \nl
0.20 & 0.80 & 0.0000 +0.1477i & 0.000 +1.704i \nl
     &      & 0.9925 -0.6113i & 11.449 -7.052i \nl
     &      & 2.0178 -0.5755i & 23.277 -6.639i \nl
0.30 & 0.70 & 0.0000 +0.0886i & 0.000 +1.303i \nl
     &      & 0.8612 -0.7196i & 12.671 -10.588i \nl
     &      & 1.7828 -0.6544i & 26.230 -9.628i \nl
0.40 & 0.60 & 0.0000 -0.0803i & 0.000 -1.391i \nl
     &      & 0.6611 -0.7167i & 11.459 -12.422i \nl
     &      & 1.6249 -0.5980i & 28.164 -10.364i \nl
0.30 & 0.30 & 0.0000 +0.1631i & 0.000 +2.409i \nl
     &      & 0.6961 -0.6842i & 10.284 -10.108i \nl
     &      & 1.6369 -0.5823i & 24.183 -8.602i \nl
0.20 & 0.30 & 0.0000 +0.2837i & 0.000 +3.309i \nl
     &      & 0.8701 -0.6169i & 10.148 -7.195i \nl
     &      & 1.8443 -0.5757i & 21.510 -6.714i \nl
0.10 & 0.30 & 0.0000 +0.1828i & 0.000 +1.238i \nl
     &      & 1.1617 -0.5938i & 7.867 -4.021i \nl
     &      & 2.4156 -0.5279i & 16.359 -3.575i \nl
\enddata
\tablecomments{Boundary conditions: $r_b=100R$, 
               $u_b=u_{\rm mff}(r_b)$}
\end{deluxetable}

\clearpage

\begin{deluxetable}{rrcc}
\footnotesize
\tablecaption{Eigenfrequencies: $l=0,1,2$ Modes}
\tablewidth{30pc}
\tablehead{
          \colhead{$\epsilon$}  & \colhead{$\mu_r$}     
         &\colhead{$\nu t_0$}  & \colhead{$\nu$ (Hz)}
}
\startdata
l = 0 &  &  &  \nl

0.20 & 0.80 & 0.0000 -0.1070i & 0.000 -0.457i \nl
     &      & 0.9243 -0.1364i & 3.951 -0.583i \nl
0.30 & 0.70 & 0.0000 -0.1607i & 0.000 -0.860i \nl
     &      & 0.9432 -0.1753i & 5.046 -0.938i \nl
0.40 & 0.60 & 0.0000 -0.2107i & 0.000 -1.317i \nl
     &      & 0.9467 -0.2193i & 5.918 -1.371i \nl

l = 1 &  &  &  \nl

0.20 & 0.80 & 0.0000 +0.3774i & 0.000 +1.613i \nl
     &      & 1.0302 -0.6324i & 4.403 -2.703i \nl
0.30 & 0.70 & 0.0000 +0.2489i & 0.000 +1.332i \nl
     &      & 1.4794 -0.5339i & 7.915 -2.857i \nl
0.40 & 0.60 & 0.0000 -0.1157i & 0.000 -0.724i \nl
     &      & 0.4794 -0.5914i & 2.997 -3.697i \nl

l = 2 &  &  &  \nl

0.20 & 0.80 & 0.0000 +0.6626i & 0.000 +2.832i \nl
     &      & 0.8896 -0.5007i & 3.803 -2.140i \nl
0.30 & 0.70 & 0.0000 +0.4764i & 0.000 +2.549i \nl
     &      & 0.8867 -0.6276i & 4.744 -3.358i \nl
0.40 & 0.60 & 0.0000 +0.1261i & 0.000 +0.789i \nl
     &      & 0.7166 -0.7765i & 4.480 -4.854i \nl
\enddata
\tablecomments{Boundary conditions: $r_b=200R$,
               $u_b=u_{\rm mff}(r_b)$}
\end{deluxetable}

\begin{deluxetable}{rrcc}
\footnotesize
\tablecaption{Eigenfrequencies: $l=0,1,2$ Modes}
\tablewidth{30pc}
\tablehead{
          \colhead{$\epsilon$}  & \colhead{$\mu_r$}     
         &\colhead{$\nu t_0$}  & \colhead{$\nu$ (Hz)}
}
\startdata
l = 0 &  &  &  \nl

0.20 & 0.80 & 0.0000 -0.1170i & 0.000 -0.130i \nl
     &      & 0.9624 -0.1318i & 1.073 -0.147i \nl
0.30 & 0.70 & 0.0000 -0.1696i & 0.000 -0.234i \nl
     &      & 0.9681 -0.1781i & 1.337 -0.246i \nl
0.40 & 0.60 & 0.0000 -0.2195i & 0.000 -0.352i \nl
     &      & 0.9682 -0.2249i & 1.554 -0.361i \nl

l = 1 &  &  &  \nl

0.20 & 0.80 & 0.0000 +1.7394i & 0.000 +1.940i \nl
     &      & 0.8352 -0.4137i & 0.931 -0.461i \nl
0.30 & 0.70 & 0.0000 +1.2636i & 0.000 +1.745i \nl
     &      & 0.9065 -0.6238i & 1.252 -0.861i \nl
0.40 & 0.60 & -0.0000 +0.3100i & -0.000 +0.497i \nl
     &      & 0.5058 -0.7329i & 0.812 -1.176i \nl

l = 2 &  &  &  \nl

0.20 & 0.80 & 0.0000 +2.7002i & 0.000 +3.011i \nl
     &      & 0.6738 -0.4998i & 0.751 -0.557i \nl
0.30 & 0.70 & 0.0000 +1.8751i & 0.000 +2.589i \nl
     &      & 0.7799 -0.5608i & 1.077 -0.774i \nl
0.40 & 0.60 & 0.0000 +0.6967i & 0.000 +1.118i \nl
     &      & 0.8149 -0.7460i & 1.308 -1.197i \nl
\enddata
\tablecomments{Boundary conditions: $r_b=500R$,
               $u_b=u_{\rm mff}(r_b)$}
\end{deluxetable}

\begin{deluxetable}{rrcc}
\footnotesize
\tablecaption{Eigenfrequencies: $l=0,1,2$ Modes}
\tablewidth{30pc}
\tablehead{
          \colhead{$\epsilon$}  & \colhead{$\mu_r$}     
         &\colhead{$\nu t_0$}  & \colhead{$\nu$ (Hz)}
}
\startdata
l = 0 &  &  &  \nl

0.20 & 0.80 & 0.0000 -0.1211i & 0.000 -0.048i \nl
     &      & 0.9768 -0.1309i & 0.390 -0.052i \nl
0.30 & 0.70 & 0.0000 -0.1736i & 0.000 -0.086i \nl
     &      & 0.9787 -0.1794i & 0.482 -0.088i \nl
0.40 & 0.60 & 0.0000 -0.2237i & 0.000 -0.128i \nl
     &      & 0.9782 -0.2275i & 0.559 -0.130i \nl

l = 1 &  &  &  \nl

0.20 & 0.80 & 0.0000 +4.9035i & 0.000 +1.958i \nl
     &      & 0.7164 -0.4018i & 0.286 -0.160i \nl
0.30 & 0.70 & 0.0000 +3.4983i & 0.000 +1.723i \nl
     &      & 0.7941 -0.5306i & 0.391 -0.261i \nl
0.40 & 0.60 & 0.0000 +1.0343i & 0.000 +0.591i \nl
     &      & 0.5838 -0.8386i & 0.333 -0.479i \nl

l = 2 &  &  &  \nl

0.20 & 0.80 & 0.0000 +7.6651i & 0.000 +3.060i \nl
     &      & 0.4746 -0.6761i & 0.189 -0.270i \nl
0.30 & 0.70 & 0.0000 +5.1606i & 0.000 +2.541i \nl
     &      & 0.6839 -0.5865i & 0.337 -0.289i \nl
0.40 & 0.60 & 0.0000 +1.8769i & 0.000 +1.072i \nl
     &      & 0.7854 -0.6765i & 0.449 -0.386i \nl
\enddata
\tablecomments{Boundary conditions: $r_b=1000R$,
               $u_b=u_{\rm mff}(r_b)$}
\end{deluxetable}

\clearpage

\begin{deluxetable}{rrcc}
\footnotesize
\tablecaption{General Relativistic and Newtonian Eigenfrequencies}
\tablewidth{30pc}
\tablehead{
          \colhead{$\epsilon$}  & \colhead{$\mu_r$}     
         &\colhead{GR $\nu$ (Hz)}  & \colhead{Newtonian $\nu$ (Hz)}
}
\startdata
l = 0 &  &  &  \nl

0.10 & 0.90 & 0.000 -0.212i & 0.000 -0.367i \nl
     &      & 5.768 -1.595i & 7.109 -0.931i \nl
0.20 & 0.80 & 0.000 -1.076i & 0.000 -1.245i \nl
     &      & 10.11 -1.722i & 11.07 -1.639i \nl
0.30 & 0.70 & 0.000 -2.195i & 0.000 -2.333i \nl
     &      & 13.39 -2.548i & 13.98 -2.666i \nl

l = 1 &  &  &  \nl

0.10 & 0.90 & 0.000 -14.26i & 0.000 +1.395i \nl
     &      & 3.174 -4.771i & 8.536 -3.167i \nl
     &      & 14.54 -4.308i & 8.536 -3.167i \nl
0.20 & 0.80 & 0.000 -1.277i & 0.000 -0.129i \nl
     &      & 6.794 -10.23i & 0.000 -9.052i \nl
     &      & 19.00 -5.833i & 17.76 -5.478i \nl
0.30 & 0.70 & 0.000 -1.976i & no nonoscillatory modes\nl
     &      & 7.120 -9.481i & 5.235 -5.264i \nl
     &      & 22.91 -6.254i & 23.22 -6.146i \nl

l = 2 &  &  &  \nl

0.10 & 0.90 & 0.000 +0.013i & 0.000 +3.301i \nl
     &      & 8.336 -3.880i & 7.310 -2.871i \nl
0.20 & 0.80 & 0.000 +1.704i & 0.000 +4.691i \nl
     &      & 11.45 -7.052i & 11.60 -6.873i \nl
0.30 & 0.70 & 0.000 +1.303i & 0.000 +0.860i \nl
     &      & 12.67 -10.59i & 21.33 -11.20i \nl
\enddata
\tablecomments{Boundary conditions: $r_b=100R$,
               $u_b=u_{\rm mff}(r_b)$}
\end{deluxetable}

\clearpage

\end{document}